\documentclass[conference,a4paper]{IEEEtran}
\usepackage{cite}
\usepackage{epsfig,graphics,mathrsfs,amssymb,amsmath,bm,color,yfonts,txfonts,multirow,multicol,slashbox,dblfloatfix}
\usepackage{subfigure}
\usepackage{algorithm2e}
\usepackage[noend]{algpseudocode}
\usepackage{flushend}
\usepackage{psfrag}

\begin{document}

\title{Carrier Aggregation in Multi-Beam High Throughput Satellite Systems}
%\author{Mirza Golam Kibria, Eva Lagunas, Nicola Maturo, Vahid Joroughi, Danilo Spano, Stefano Andrenacci,\\Symeon Chatzinotas and Björn Ottersten
\author{Mirza Golam Kibria, Eva Lagunas, Nicola Maturo,  Danilo Spano, Hayder Al-Hraishawi and Symeon Chatzinotas\\
SIGCOM Research Group, SnT, University of Luxembourg\\
%E-mails:\{mirza.kibria, eva.lagunas, nicola.maturo, danilo.spano, hayder.al-hraishawi, symeon.chatzinotas\}@uni.lu
\thanks{The authors are with SIGCOM Research Group, SnT, University of Luxembourg (e-mails: $\left\{\text{mirza.kibria}\right\}@\text{uni.lu}$). }
}
\maketitle

\begin{abstract}
%\vspace{-2mm}
Carrier Aggregation (CA) is an integral part of current terrestrial networks. Its ability to enhance the peak data rate, to efficiently utilize the limited available spectrum resources and to satisfy the demand for data-hungry applications has drawn large attention from different wireless network communities. Given the benefits of CA in the terrestrial wireless environment, it is of great interest to analyze and evaluate the potential impact of CA in the satellite domain. In this paper, we study CA in multibeam high throughput satellite systems. We consider both inter-transponder and intra-transponder CA at the satellite payload level of the communication stack, and we address the problem of carrier-user assignment assuming that multiple users can be multiplexed in each carrier. The transmission parameters of different carriers are generated considering the transmission characteristics of carriers in different transponders. In particular, we propose a flexible carrier allocation approach for a CA-enabled multibeam satellite system targeting a proportionally fair user demand satisfaction.
Simulation results and analysis shed some light on this rather unexplored scenario and demonstrate the feasibility of the CA in satellite communication systems.
\end{abstract}

%\begin{keywords}
%High Throughput Satellite, Carrier Aggregation, Flexible Resource Allocation.
%\end{keywords}
\IEEEpeerreviewmaketitle

\section{Introduction}

During the last decade, satellite technology has been rapidly growing due to the immense benefits that satellite communication systems can provide, such as ubiquitous broadband coverage over a large area, wideband transmission capability, and navigation assistance\cite{Sharma}. Because of these benefits, the satellite data traffic is witnessing a phenomenal growth contributed by the delivered telecommunication services in a wide range of sectors such as aeronautical, maritime, military, rescue and disaster relief\cite{Vasavada}. Moreover, the unprecedented number of emerging applications such as high definition television, interactive multimedia services and broadband internet access is leading to an escalating need of flexible satellite systems\cite{Sharma}, where the available resources have to be dynamically assigned according to the traffic demands.

With the ever-increasing satellite communication traffic and the rapidly growing demands for anytime, and anywhere access to satellite services, the satellite spectrum resources need to be efficiently and thoroughly utilized because the system capacity significantly depends on available satellite resources and their utilization. Few studies have been conducted from different perspectives for the purpose of enhancing satellite system capacity. For instance, in \cite{Knab2013}, the satellite transponder power and the required terminal power for a group of terminals on the transponder have been optimized with taking into consideration the throughput-power trade-off. Reference \cite{Cocco2015} investigates radio resource allocation in the forward link of multibeam satellite networks and develops an allocation algorithm  to meet the requested traffic across the different beams while taking fairness into account. The essential satellite system parameters such as uplink and downlink satellite antenna gains, the ground terminals' receive gain and noise temperature, path losses and fades, data rates have been jointly optimized in \cite{Knab2015} to improve resource utilization.

On a parallel avenue, the concept of carrier aggregation (CA) emerged as a promising technology allowing the mobile terrestrial network operators to combine multiple component carriers across the available spectrum in order to extend the channel bandwidth, and hence, increasing the network data throughput and overall capacity \cite{3GPP-CA,3GPP-CA1}. Enabling CA feature in cellular network attains significant gains in performance through exploiting the available spectrum resources and satisfying the high throughput demands. Interestingly, CA does not only address the spectrum scarcity and boost capacity fairness among the users but also maintains the system quality of service via efficient interference management and avoidance capabilities \cite{CA2}. While the application CA in terrestrial scenarios has
been widely considered, its application in satellite communications is still a rather unexplored area. Recently, the application of CA in satellite communications has received interest in an European Space Agency (ESA) funded project named CADSAT\cite{ESA}, where several potential scenarios have been discussed and
analyzed based on market, business and technical feasibility. In this paper, we
focus on one of the preselected scenarios which is the multibeam multicarrier geostationary earth orbit (GEO) satellite system.

Channel bonding as defined in DVB-S2X standard\cite{DVBS2X} is in many ways similar to the concept of CA. CA refers to aggregate multiple contiguous and non-contiguous carriers in different spectrum bands, and then, can be used simultaneously, whereas, channel bonding combines multiple adjacent channels to constitute larger transmission bandwidths \cite{NewTec,ZKhan}. However, channel bonding as defined in DVB-S2X standard has several inherent limitations for broadband applications that might restrict the essential resource allocation flexibility. For example, channel bonding is mainly focusing on aggregating carriers across transponders where the maximum number of bonded transponders is three. Moreover, the bonded channels has to be located in the same frequency band. Channel bonding employs constant coding and modulation schemes, where all the services undergo the same coding and modulation procedure, which is a very obstructive factor for its employment in the emerging broadband applications. Having been motivated by these facts, this paper is considering CA to circumvent these limitations and improve system flexibility.

\begin{figure*}
  \centering
   \includegraphics[scale=.46]{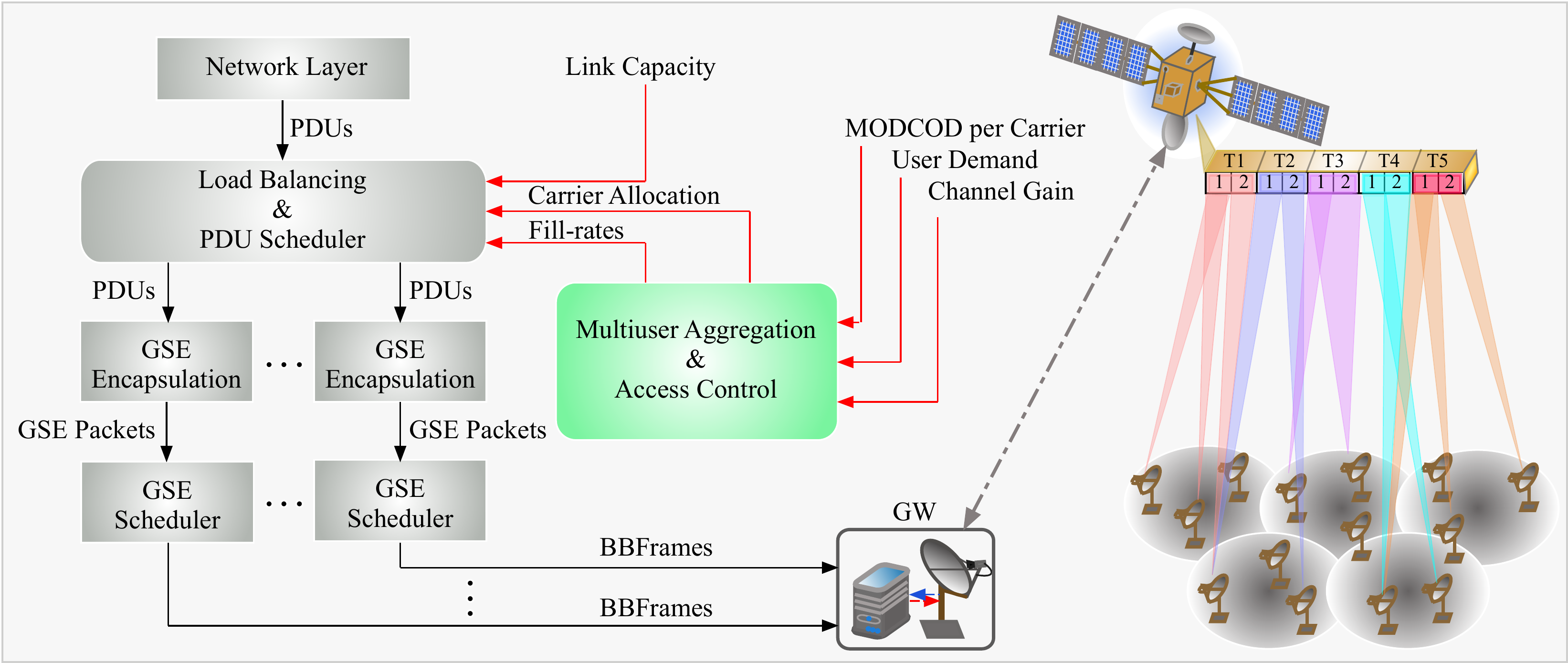}
   \caption{Schematic model of the CA. In this example, there are 5 multicarrier transponders, namely $ T_1, T_2, \cdots, T_5$, each with 2 component carriers. There can be inter-transponder CA as seen in $ T_1$ and $ T_2$ where user of $T_2$ is served by carriers of $T_1$ and $T_2$ as well as intra-transponder CA as seen in $ T_3$ where both carriers served one of its users.}
   \label{fig_CADSAT_1}
\end{figure*}

\noindent
\textbf{Contributions:} The main technical contributions of this paper can be summarized as follows:
\begin{enumerate}
	\item Adopting CA techniques in satellite mobile communication systems has been investigated, and the effects of intra-transponder and inter-transponder CA at payload level of the communication stack have been thoroughly analyzed.
	\item An efficient multi-user (MU) aggregation scheme for CA considering user achievable capacities over different carriers has been proposed. In particular, the user-carrier association and optimal carrier fill-rates are obtained, where fill-rate defines the percentage of carrier bandwidth being assigned to a given user.
	\item The performance of the proposed CA solution has been evaluated based on its capability in minimizing the unmet and unused capacity. Simulation results are provided to confirms the efficacy of the proposed solution.
\end{enumerate}

The paper is organized as follows. Section II provides the system model of this study. In section III, we present the multiuser aggregation and access control in the CA problem statement and the proposed solution. Section IV presents the simulation results and section V draws conclusions.

\textit{Notations}: Boldface lower-case and upper-case letters define vectors and matrices, respectively. $\mathbb{R}$ defines a real space. Superscript $(\cdot)^T$ denotes the transpose oprtation and $\rm{diag}(\cdot)$ puts the diagonal elements of a matrix into a vector. Operator  $\text{vec}(\cdot)$ stacks all the elements of the argument into a vector and and $||\cdot||_1$ returns 1-norm of the argument. ${\bf 1}^{x\times y}$ defines a vector/matrix of all one elements, and $\bf 1^X$ denotes a matrix of ones with dimensions same as of  matrix $\bf X$. ${\bf 0}^{x\times y}$ defines a vector/matrix of all zero elements.

\section{System Model}
We consider a multi-beam GEO satellite system that employs multi-carrier transponders. In particular, we consider the intra-satellite CA scenario, where both intra-transponder and inter-transponder CA take place. Let the total number of users in the system be $N_{\rm U}$ while the total number of carriers is $N_{\rm C}$. Each carrier has a bandwidth of $B_{\rm w}$ MHz.  The number of users and carriers may vary among the beams. The users are classified into two service level agreement (SLA) groups, such as premium users and non-premium users. The premium users allowed to aggregate up to $\Delta_{\max}$ component carriers depending on their demand while the non-premium users operate always in a single carrier mode. The carriers may be shared with multiple users, independent of the group they belong depending on the SLAs. The schematic model of our considered system is given in Fig.~\ref{fig_CADSAT_1}, where the satellite has five multi-carrier transponders each with two carriers.

In our considered system model the carrier assignment is dynamic based on user traffic demand. Based on the user demand and link budget per carrier, the user-carrier association is determined along with the fill-rates. Instead of allowing the premium users be constantly logged-on in the two or more carriers (even if they are not using them all the time), we rather consider that the carriers are dynamically enabled to premium terminals when needed. However, this mode implies more complexity in terms of user traffic monitoring and reconfigurability of the system. 
The operations pertaining to the network, MAC/link layer of the communication stack, i.e., load balancing, packet data unit (PDU) scheduling, generic stream encapsulation (GSE), GSE packet scheduling over the baseband frames\cite{DVBS2X} for a given CA user are also depicted in  Fig.~\ref{fig_CADSAT_1}.  
Due to space limitation, in this paper we focus on the MU aggregation and access control block design.

\section {Multiuser Aggregation and Access Control}
In this section, we present the solution we propose for MU aggregation and access control for efficient CA operation in multibeam satellite systems. As mentioned earlier that the carriers may be shared with multiple users, independent of the group they belong depending on the SLAs, let us now define below some important parameters that we make use of  
\begin{itemize}
\item $a_{c,u}$: association indicator: if carrier $c$ is a component carrier of user $u$, then $a_{c,u}=1$, otherwise 0. Let us store all $a_{c,u}$ ($c=1,2,\cdots,N_{\rm C}$ and $u=1,2,\cdots,N_{\rm U}$) in $\bf A$. Therefore, $\bf A$ (association matrix) with $a_{c,u}\in\{0,1\}$ is a binary matrix of size $N_{\rm C}\times N_{\rm U}$.

\item $f_{c,u}$: fill-rate variable. A carrier may be shared by multiple users.  The value of $f_{c,u}$ lies between 0 and 1 as we use normalized value of the percentage of carrier $c$'s bandwidth being assigned to user $u$. If carrier $c$ is used in part by user $u$, $f_{c,u}$ will be $>0$.  Let us store all $f_{c,u}$ ($c=1,2,\cdots,N_{\rm C}$ and $u=1,2,\cdots,N_{\rm U}$) in $\bf F$. Therefore, $\bf F$ (fill-rate matrix)  of size $N_{\rm C}\times N_{\rm U}$ is a positive matrix with elements $0\le f_{c,u}\le 1$.

\item $r_{c,u}$: achiavable rate value. It defines the rate achievable by user $u$ if the component carrier $c$ is assigned to user $u$ assuming $f_{c,u}=1$. Let us store all $r_{c,u}$ ($c=1,2,\cdots,N_{\rm C}$ and $u=1,2,\cdots,N_{\rm U}$) in $\bf R$. Therefore, $\bf R$ (achievable rate matrix) is a positive matrix of size $N_{\rm C}\times N_{\rm U}$. It is a knwon matrix as the achievable rate can be calculated based on channel-state information and link budget.
\end{itemize}

The MU aggregation and access control is constrained by ($i$) the maximum number of carriers that can be aggregated by a signle premium user, which depends on the decoding capability of the user terminal chipset, ($ii$) the summation of fill-rates of different users for any given carrier must not exceed 100\%, i.e., $\sum_{u=1}^{N_{\rm U}}f_{c,u}\le 1$, and ($iii$) adaptation of carrier assignment problem to the dynamic variations of demand with minimal amount of carrier swapping to reduce the signaling overhead and link outage/degradation. Let us assume that we have a system running with a given ${\bf A}_t$ (${\bf A}$ at time-instant $t$) and the demands change significantly over time. Then we need to update ${\bf A}_t$ to ${\bf A}_{t+1}$. Ideally, we would prefer to move as fewer users as possible to minimize signaling overhead and link outage/degradation during the carrier swapping. As demand changes, and we need to re-design  $\bf A$ and $\bf F$ such that the difference from the previous state is minimal. We can write this constraint as $||\text{vec}({\bf A}_{t+1})-\text{vec} ({\bf A}_{t})||_1\le Q$, where $Q$ is the maximum number of changes allowed in the subsequent carrier assignment. As a results, we have the following constraints in our CA optimization problem based on the constraints ($i$), ($ii$) and ($iii$).
\begin{equation}
\label{main5555}
 \begin{array}{*{35}{l}}
\sum_{c=1}^{N_{\rm C}} a_{c,u}\le  \Delta_{\max}, u=1,2,\cdots, N_{\rm U}\\
\sum_{u=1}^{N_{\rm U}} f_{c,u}\le 1, c=1,2,\cdots, N_{\rm C}\\
||\text{vec}({\bf A}_{t+1})-\text{vec} ({\bf A}_{t})||_1\le Q\\
\end{array}
\end{equation} 
\noindent where $ \Delta_{\max}$ the the maximum number of parallel streams, i.e., carriers the user terminal chipset can decode simultaneously.

Let $d_u$ be the demand of user $u$.
The offered capacity to user $u$ is calculated by $s_u=\sum_{c=1}^{N_{\rm C}}a_{c,u}f_{c,u}r_{c,u}$. 
The MU aggregation and access control optimization problem in CA is formulated to maximize the minimum ratio between the offered capacity and the requested demand. Based on our system model and the constraints already discussed, the problem can be expressed as follows
\begin{equation}
\label{main123}
 \begin{array}{*{35}{l}}
\underset{a_{c,u},f_{c,u}}{\max}\hspace{1mm}\underset{u}{\min}\hspace{1mm}\frac{s_u}{d_u}\\
\text{}\text{subject to }\text{ C1:} \hspace{2mm}s_u=\sum_{c=1}^{N_{\rm C}}a_{c,u}f_{c,u}r_{c,u},\vspace{1.5mm} \\
\text{}\hspace{15mm}\text{ C3:} \hspace{2mm}\sum_{c=1}^{N_{\rm C}} a_{c,u}\le \Delta_{\max}, u=1,2,\cdots, N_{\rm U}, \vspace{1.5mm} \\
\text{}\hspace{15mm}\text{ C4:} \hspace{2mm}\sum_{u=1}^{N_{\rm U}} f_{c,u}\le 1, c=1,2,\cdots, N_{\rm C}, \vspace{1.5mm} \\
\text{}\hspace{15mm}\text{ C5:} \hspace{2mm}a_{c,u}\in\{0,1\},  u=1,\cdots, N_{\rm U}, c=1,\cdots, N_{\rm C} \\
\text{}\hspace{15mm}\text{ C6:} \hspace{2mm}0\le f_{c,u}\le 1,  u=1,\cdots, N_{\rm U}, c=1,\cdots, N_{\rm C} \\
\text{}\hspace{15mm}\text{ C7:} \hspace{2mm} ||\text{vec}({\bf A}_{t+1})-\text{vec} ({\bf A}_{t})||_1\le Q
\end{array}
\end{equation}
We can simplify the $\max-\min$ optimization problem by turning it into a maximization problem with the help of an additional slack variable $\psi$ along with a new constraint $\frac{s_u}{d_u}\ge \psi, \text{i.e., }s_u\ge \psi d_u$.
The optimization problem in \eqref{main123} is a mixed-integer non-linear programming problem as we have the non-linear constraint $s_u=\sum_{c=1}^{N_{\rm C}}a_{c,u}f_{c,u}r_{c,u}$ as well as binary integer variables, which is computationally very expensive. We propose an efficient solution to this problem. 

%\vspace{-6mm}
\subsection{Proposed Solution}
Note that $a_{c,u}\in\{0,1\}$ is a binary variable while $0\le f_{c,u}\le1$ is a continuous variable, and we have their multiplication in constraint C1, which is non-linear. We employ the following technique to transform the constraint into linear constraint.
Here, $a_{c,u}$ (binary integer ) and $f_{c,u}$ (continuous) are the optimization variables in \eqref{main123}, and we need to deal with their product $a_{c,u}f_{c,u}$. Note that if both $a_{c,u}$ and $f_{c,u}$ were continuous variables, we would have ended up having a quadratic nonlinear programming problem instead of mixed-integer non-linear programming problem. When  quadratic terms appear in constraints, it creates issues with convexity. However, constraint C1 is distinctive as $a_{c,u}$ is a binary variable and $f_{c,u}$ is a bounded continuous variable. The nonlinear term or the product $a_{c,u}f_{c,u}$ can be linearized by introducing auxiliary variables ${\lambda}_{c,u}=a_{c,u}f_{c,u}$ and incorporating the following linear constraints into the optimization problem. 
\vspace{-2mm}
\begin{equation}
\label{main5}
 \begin{array}{*{35}{l}}
{\rm{min}}\hspace{1mm}\{0,f_{c,u}^{\rm{lb}}\}\le {\lambda}_{c,u} \le {\rm{max}} \hspace{1mm} \{0,f_{c,u}^{\rm{up}}\}\vspace{3mm} \\
f_{c,u}^{\rm{lb}}a_{c,u} \le {\lambda}_{c,u} \le f_{c,u}^{\rm{ub}}a_{c,u} \vspace{3mm} \\
f_{c,u}-f_{c,u}^{\rm{ub}}(1-a_{c,u})\le {\lambda}_{c,u} \le f_{c,u}-f_{c,u}^{\rm{lb}}(1-a_{c,u}) 
\end{array}
\end{equation} 
Here, $f_{c,u}^{\rm{lb}}$ and $f_{c,u}^{\rm{ub}}$ are the lower bound and upper bound, respectively, of the continuous variable $f_{c,u}$.
Accoding to the definition of $f_{c,u}$, we have  $f_{c,u}^{\rm{lb}}=0$ and  $f_{c,u}^{\rm{ub}}=1$. After the linearization of nonlinear constraint C1, the following linear constraints are assimilated in \eqref{main123}, and constraint C1 now becomes $s_u=\sum_{c=1}^{N_{\rm C}}\lambda_{c,u}r_{c,u}$, which is linear.
\begin{equation}
\label{main6}
\text{C8:} \left\{ \begin{array}{*{35}{l}}
\text{}\hspace{1mm}\text{N1:} \hspace{2mm}{\lambda}_{c,u}\le a_{c,u},\hspace{20mm}\forall c,u \vspace{2mm} \\
\text{}\hspace{1mm}\text{N2:}\hspace{2mm}{\lambda}_{c,u}\ge 0, \hspace{23.4mm}\forall c,u \vspace{2mm} \\
\text{}\hspace{1mm}\text{N3:}\hspace{2mm} {\lambda}_{c,u}\le f_{c,u}, \hspace{18.5mm}\hspace{2mm}\forall c,u\vspace{2mm} \\
\text{}\hspace{1mm}\text{N4:}\hspace{2mm} {\lambda}_{c,u}\ge f_{c,u}-(1-a_{c,u}),\hspace{3.0mm}\hspace{1.2mm}\forall c,u 
\end{array}\right.
\end{equation}
%\vspace{-10mm}
For the case, $a_{c,u}=0$, $\lambda_{c,u}$ or the product ${\lambda}_{c,u}=a_{c,u}f_{c,u}$ should be 0. The inequalities \{N1, N2\} causes $0\le {\lambda}_{c,u}\le 0$, yielding ${\lambda}_{c,u}$ to be 0. The other pair of linear constraints \{N3, N4\} returns $f_{c,u}-1\le {\lambda}_{c,u} \le f_{c,u}$, and ${\lambda}_{c,u}=0$ conforms these inequalities. On the other hand, for the case $a_{c,u}=1$, the product should be ${\lambda}_{c,u}=f_{c,u}$. The inequalities N1 and N2 enforce $0 \le {\lambda}_{c,u} \le 1$, which is satisfied by ${\lambda}_{c,u}=f_{c,u}$. The second pair of inequalities N3 and N4 yields $f_{c,u} \le {\lambda}_{c,u} \le f_{c,u}$, forcing ${\lambda}_{c,u}=f_{c,u}$ as needed. This linearization approach, in principle, splits the feasible regions into two subregions, one when $a_{c,u}=0$ and $f(a_{c,u},f_{c,u})=a_{c,u}f_{c,u}=0$ (trivially linear) and the other when $a_{c,u}=1$ and $f(a_{c,u},f_{c,u})=f_{c,u}$ (also linear). After the linearization, the problem in \eqref{main123} becomes a mixed-integer linear programming problem which is given below
%\vspace{-18mm}
\begin{equation}
\label{main124}
 \begin{array}{*{35}{l}}
\hspace{9.5mm}\underset{a_{c,u},f_{c,u},\lambda_{c,u}}{\max} \hspace{1mm}\psi\\
\text{}\text{subject to }\text{ C1:} \hspace{2mm}s_u=\sum_{c=1}^{N_{\rm C}}\lambda_{c,u}r_{c,u},\vspace{1.5mm} \\
\text{}\hspace{15mm}\text{ C2:} \hspace{2mm}s_u\ge \psi d_u,\\
\text{}\hspace{15mm}\text{ C3:} \hspace{2mm}\sum_{c=1}^{N_{\rm C}} a_{c,u}\le \Delta_{\max},  u=1,2,\cdots, N_{\rm U}, \vspace{1.5mm} \\
%\text{}\hspace{16.5mm}0\le\psi_i\le S_{\rm max} . \vspace{1.5mm} \\
\text{}\hspace{15mm}\text{ C4:} \hspace{2mm}\sum_{u=1}^{N_{\rm U}} f_{c,u}\le 1,  c=1,2,\cdots, N_{\rm C}, \vspace{1.5mm} \\
\text{}\hspace{15mm}\text{ C5:} \hspace{2mm}a_{c,u}\in\{0,1\}, \vspace{1.5mm} \\
\text{}\hspace{15mm}\text{ C6:} \hspace{2mm}0\le f_{c,u}\le 1, \vspace{1.5mm} \\
\text{}\hspace{15mm}\text{ C7:}||\text{vec}({\bf A}_{t+1})-\text{vec} ({\bf A}_{t})||_1\le Q\\
\hspace{16mm}\text{C8:} \left\{ \begin{array}{*{35}{l}}
\text{}\hspace{1mm}\text{N1:} \hspace{2mm}{\lambda}_{c,u}\le a_{c,u},\hspace{20mm} \\
\text{}\hspace{1mm}\text{N2:}\hspace{2mm}{\lambda}_{c,u}\ge 0, \hspace{30.4mm} \\
\text{}\hspace{1mm}\text{N3:}\hspace{2mm} {\lambda}_{c,u}\le f_{c,u}, \hspace{24.5mm}\hspace{2mm}\\
\text{}\hspace{1mm}\text{N4:}\hspace{2mm} {\lambda}_{c,u}\ge f_{c,u}-(1-a_{c,u}),\hspace{1.0mm}\hspace{1.2mm}
\end{array}\right.
\end{array}
\end{equation}
Like $\bf A$ and $\bf F$, we can store all $\lambda_{c,u}$ in matrix $\bf \Lambda$ of size $N_{\rm C}\times N_{\rm U}$. When $s_u$'s are stored in a vector ${\bf s}\in\mathbb{R}^{1\times N_{\rm U}}$, we can express $\bf s$ as $\text{diag} ({\bf \Lambda}^T {\bf R})$. Let ${\bf d}\in\mathbb{R}^{1\times N_{\rm U}}\triangleq {{\bf d}}=[d_1,d_2,\cdots, d_{N_{\rm U}}]$. Similarly, the constraints in (i) and (ii) can be expressed as ${\bf a}\triangleq {\bf \left(1^A\right)}^T{\bf A}\le \Delta_{\rm max}{\bf 1}^{1\times N_{\rm U}}$ and ${\bf f}\triangleq {\bf \left(1^{\left(F^T\right)}\right)}^T{\bf F}^T\le {\bf 1}^{1\times N_{\rm C}}$, respectively. Therefore, we can also express \eqref{main124} as

\begin{equation}
\label{main125}
 \begin{array}{*{35}{l}}
\hspace{9.5mm}\underset{\bf A,F,\Lambda}{\max} \hspace{1mm}\psi\\
\text{}\text{subject to }\text{ C1:} \hspace{2mm}{\bf s}=\text{diag} ({\bf \Lambda}^T {\bf R}),\vspace{1.5mm} \\
\text{}\hspace{15mm}\text{ C2:} \hspace{2mm}{\bf s}\succeq \psi{\bf d},\\
\text{}\hspace{15mm}\text{ C3:} \hspace{2mm}{\bf a}\preceq \Delta_{\max}{\bf 1}^{1\times N_{\rm U}}, \vspace{1.5mm} \\
\text{}\hspace{15mm}\text{ C4:} \hspace{2mm}{\bf f}\preceq {\bf 1}^{1\times N_{\rm C}}, \vspace{1.5mm} \\
\text{}\hspace{15mm}\text{ C5:} \hspace{2mm}{\bf A}\in\{0,1\}, \vspace{1.5mm} \\
\text{}\hspace{15mm}\text{ C6:} \hspace{2mm}{\bf 0}^{N_{\rm C}\times N_{\rm U}}\preceq{\bf F}\preceq {\bf 1}^{N_{\rm C}\times N_{\rm U}}, \vspace{1.5mm} \\
\text{}\hspace{15mm}\text{ C7:}||\text{vec}({\bf A}_{t+1})-\text{vec} ({\bf A}_{t})||_1\le Q\\
\hspace{16mm}\text{C8:} \left\{ \begin{array}{*{35}{l}}
\text{}\hspace{1mm}\text{N1:} \hspace{2mm}{\bf  \Lambda}\preceq {\bf A},\hspace{20mm} \\
\text{}\hspace{1mm}\text{N2:}\hspace{2mm}{\bf  \Lambda}\succeq {\bf 0}^{N_{\rm C}\times N_{\rm U}}, \hspace{30.4mm} \\
\text{}\hspace{1mm}\text{N3:}\hspace{2mm}{\bf  \Lambda}\preceq {\bf F}, \hspace{24.5mm}\hspace{2mm}\\
\text{}\hspace{1mm}\text{N4:}\hspace{2mm}{\bf  \Lambda}\succeq{\bf F}-({\bf 1}^{N_{\rm C}\times N_{\rm U}}-{\bf A}),\hspace{1.0mm}\hspace{1.2mm}
\end{array}\right.
\end{array}
\end{equation}
Here, the relation ${\bf x}\succeq {\bf y}$ states that an element in $\bf x$ succeds the same indexed element in $\bf y$, i.e., $x_i\ge y_i$, while the relation ${\bf x}\preceq {\bf y}$ states that an element in $\bf x$ precedes the same indexed element in $\bf y$, i.e., $x_i\le y_i$. 
The optimization problem in \eqref{main125} is a mixed-integer linear programming problem and can be efficiently solved by optimization toolbox like CVX\cite{CVX}.

\section{Simulation Results}

The simulation set-up for evaluating the performance of CA in high throughput satellite system is as follows.  A 71-beam GEO satellite beam pattern provided by ESA is considered. In this framework, we extract a cluster of 8 adjacent beams from the total pattern. The number of users in each beam ranges from 30 to 35, which are randomly distributed over the coverage of the  extracted cluster. Each beam has two carriers, and the carrier bandwidth is $54$ MHz. The transmit power per beam is set to 10 Watt. 5\% of the users are taken to be very high demand users while the remaining users have low/average demand. The simulation parameters are provided in Table.~\ref{tab-123}. The proposed CA scheme is evaluated by quantifying peak and the average rate of the users to assess the gains with respect to the system without CA.

 \begin{table}[h]
\centering
\caption{Simulation Parameters} 
\label{tab-123}
\begin{tabular}{l |r}
\hline
\hline
Satellite longitude & \hspace{10mm}$30^{\circ}{\rm E}$ (GEO)\\ 
Number of carriers per beam, & 2\\
Transmit power per beam, $P_{\rm T}$ &\hspace{2mm}10 W\\
Number of beams, $N_{\rm B}$ & \hspace{2mm}8\\
Beam radiation pattern & \hspace{2mm}Provided by ESA\\ 
Downlink carrier frequency & \hspace{2mm}19.5 GHz\\ 
Carrier bandwidth, $B_{\rm W}$ & \hspace{1.95mm}54 MHz \\ 
Roll-off factor & \hspace{1.95mm}20\% \\ 
Maximum number of decoded carriers, $\Delta_{\max}$ & 2\\
\hline
\end{tabular}
\end{table}

One of the figure of merits for resource allocation in satellite communications is the unmet capacity, which is the total amount of demanded capacity that cannot be satisfied with the available resources. The unmet capacity is defined as $C_{\rm unmet}=\sum_{i=1}^{N_{\rm U}}(d_u-s_u)^+$,  where $(x)^+=\max (0,x)$. Similarly, excess capacity is another figure of merit that corresponds to the sum of offered capacity across the beams which exceeds the demanded capacity, which is defined as $C_{\rm unused}=\sum_{i=1}^{N_{\rm U}}(s_u-d_u)^+$. The  $C_{\rm unmet}$ and $C_{\rm unused}$ values deliver evidence of efficiency of the proposed CA solution.

\begin{figure}
  \centering
   \includegraphics[scale=0.56]{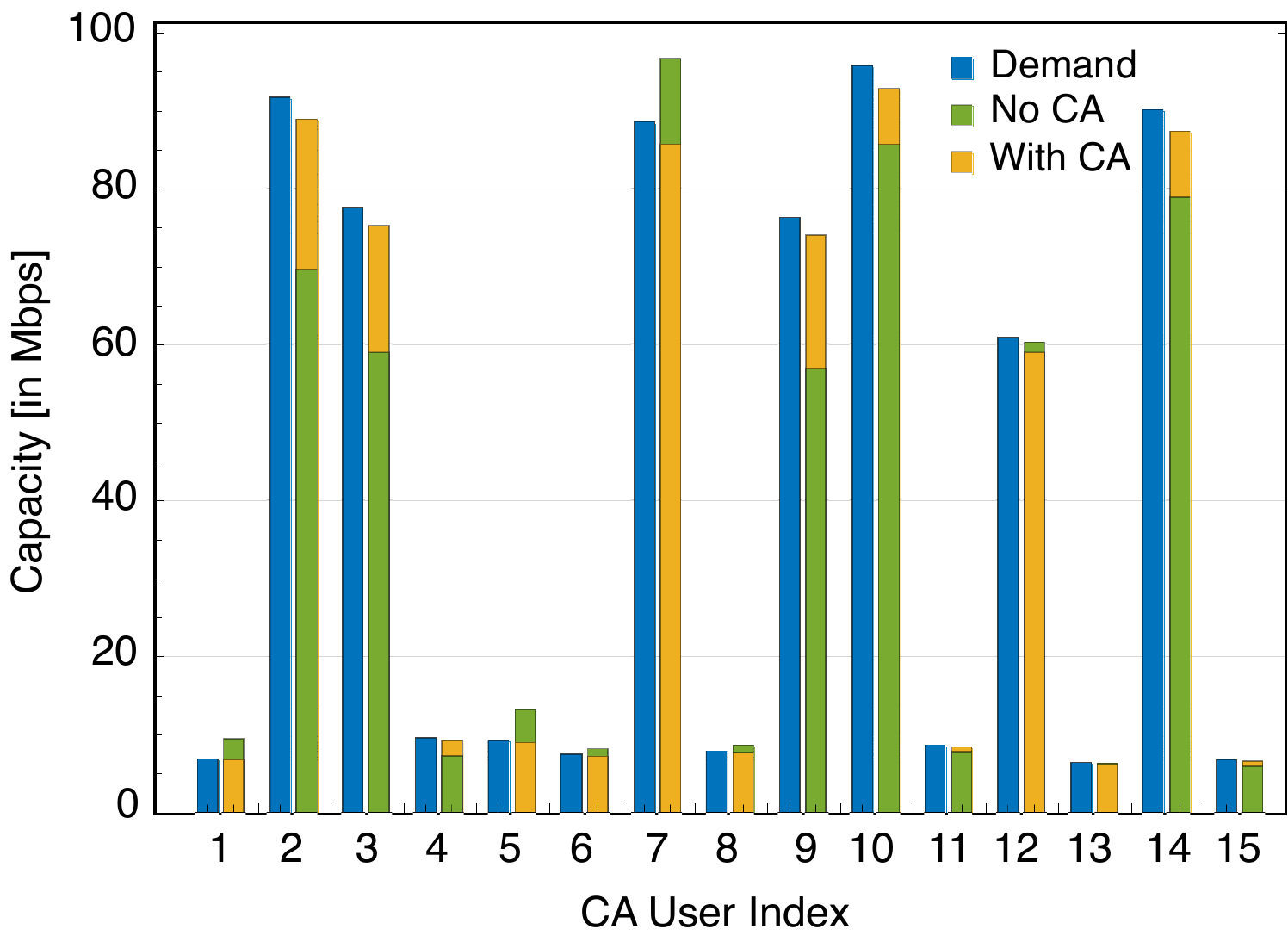}
   \caption{Achievable supply capacity with and without CA. The demand and supply values only for the CA users are depicted.}
   \label{fig_CADSAT_res1}
\end{figure}
In Fig.~\ref{fig_CADSAT_res1}, we evaluate the performance of the proposed CA solution in terms of its capability in enhancing the peak data rate of the high demand users as well as in rate-matching. The bar chart shows the performances only for CA users. In can be seen that with CA, the demands of the users are well satisfied. The yellow part on top of the blue bars reflects the additional capacity provided with CA. Although the provided capacity to some of the users by the system without CA is higher than that with CA, the rate matching is not as good as with the proposed CA solution. It is also very evident that with CA, high demand users can be satisfied. It may happen that with CA, the supply capacity can be sometimes lower than that without CA. For example, for the CA user with index 7, the supply capacity with CA is lower than the supply capacity without CA. Note that the proposed solution for CA in this study not only aims at supplying capacity as closely as possibe to the demand capacity, but also opts to treat all the users as fairly as possible. Hence, for CA users 5, 7, 12, etc., the supply capacity with CA is lower than that without CA is just becuase of the fairness feature that has been  infused in the system.

\begin{figure}
  \centering
   \includegraphics[scale=0.645]{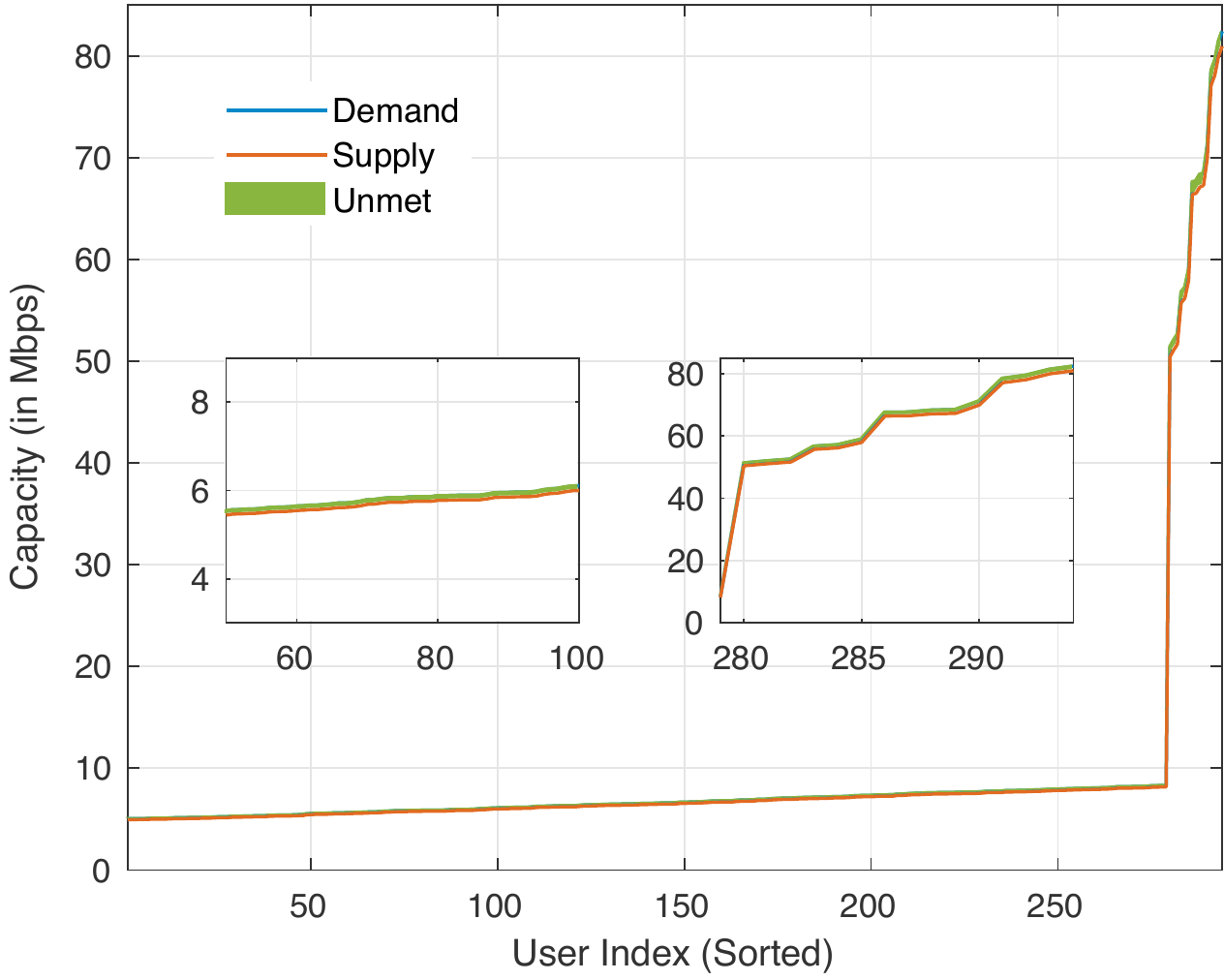}
   \caption{Unmet vs. unused capacity comparison between satellite systems without CA. Th inset plots are zoomed out depiction of some parts of the base plot. The users are sorted based on their demands.}
   \label{fig_CADSAT_res2}
\end{figure}

\begin{figure}
  \centering
   \includegraphics[scale=0.645]{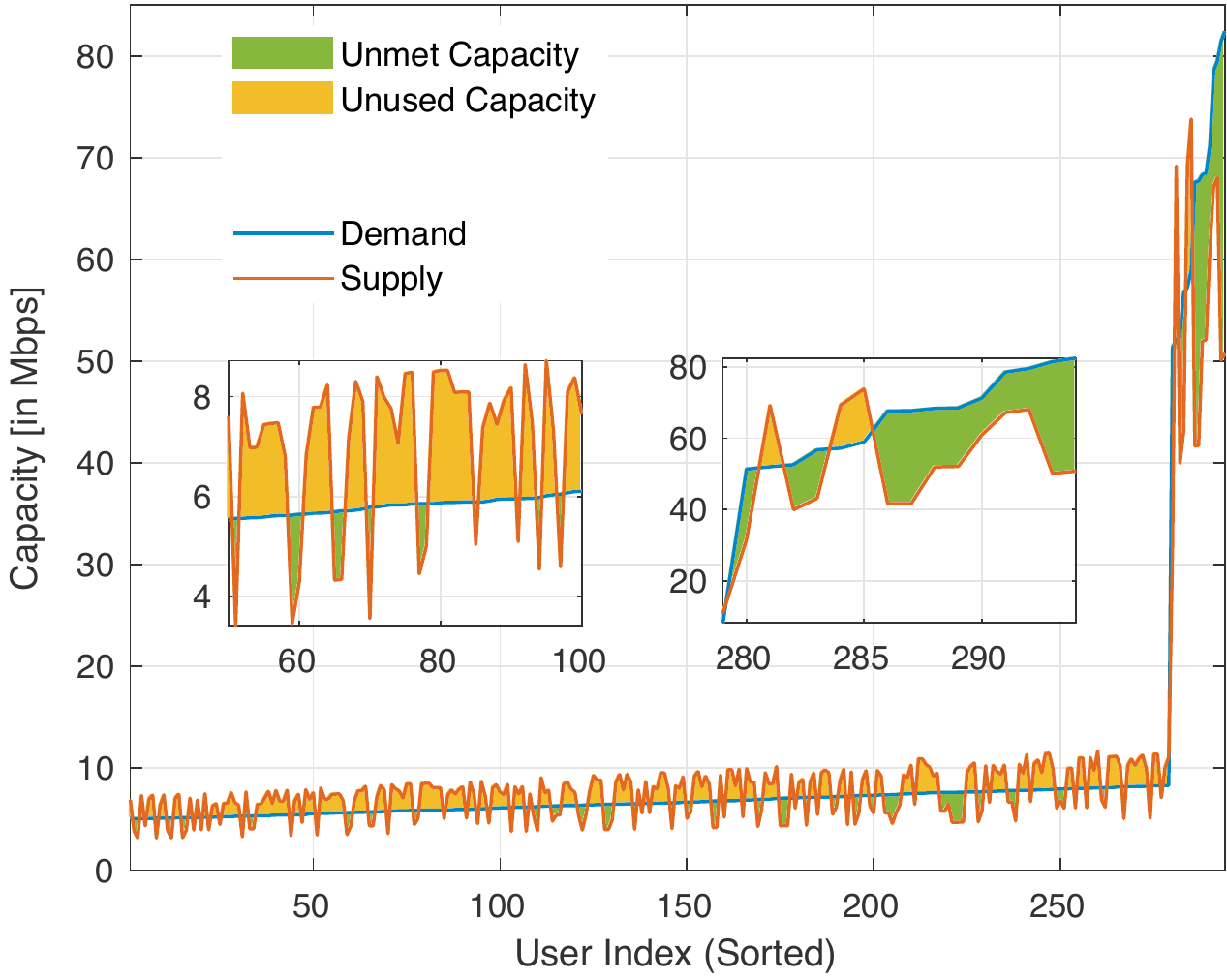}
   \caption{Unmet vs. unused capacity comparison between satellite systems with CA. Th inset plots are zoomed out depiction of some parts of the base plot. The users are sorted based on their demands.}
   \label{fig_CADSAT_res3}
\end{figure}

In Fig.~\ref{fig_CADSAT_res2} and Fig.~\ref{fig_CADSAT_res3} we evaluate and compare the systems with and without CA in terms of unmet and unused capacity.
Fig.~\ref{fig_CADSAT_res2} shows the unused and unmet capacity without CA and rate-matching while Fig.~\ref{fig_CADSAT_res3} depicts the performance with CA along with our proposed rate-matching solution. It is evident from the performances that the proposed CA solution performs exceptionally well in utilizing the satellite resources, i.e., in reducing the unmet and unused capacity. In this current evaluation, the total demand in the system is 2.837 Gbps. The supply capacity without CA is 2.873 Gbps while the supply capacity with CA is 2.787 Gbps. The unmet and unused capacity without CA are 396 Mbps and 433 Mbps, respectively. On the other hand, the unmet and unused capacity with CA are 53 Mbps and 0 Mbps, respectively. Note that in case of the system without CA, the available satellite data-rate was assigned proportionally among the users based on their demand. As the objective of our proposed MU access for CA is to maximize the rate matching, the total supply with the proposed CA is lower than that without CA. However, the amount of unmet and unused capacity with the proposed solution is much smaller than that without CA. The inset figures in Fig.~\ref{fig_CADSAT_res2} exhibit that without CA, the unused capacity is higher for the low demand users while the high demand users have a relatively higher unmet capacity. While with CA, the unused/unmet capacity remains very low for all the users. Note that for the simulation results so far, the constraint C7 in \eqref{main124} has been ignored as we just evaluate the performance of the proposed CA solution for one particular demand profile.

\begin{figure}
  \centering
   \includegraphics[scale=0.655]{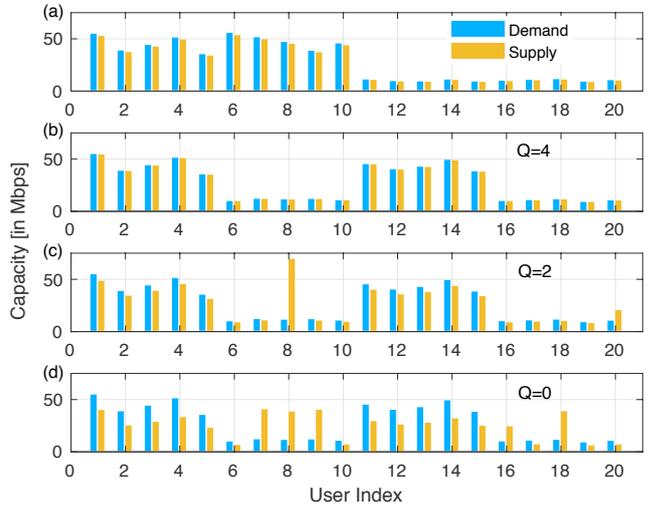}
   \caption{Impact of values of $Q$ on achievable capacity and rate matching charateristics.}
   \label{fig_CADSAT_res4}
%\vskip3em
 \psfrag{Demand}{$Demand$}
\end{figure}
%\vspace{-5mm}

Fig.~\ref{fig_CADSAT_res4} reflects how the proposed CA solution reacts to changes in user demands depending on different values of $Q$ in C7 of \eqref{main125}. We consider two different demand profiles for the users in the system. Fig.~\ref{fig_CADSAT_res4}(a) belongs to demand profile 1 and the remaining subplots ((b) to (d)) in Fig.~\ref{fig_CADSAT_res4} belong to demand profile 2. Here, we consider a system of 2 beams extracted from the 71-beam pattern and the beams have 20 users each. Under demand profile 1, users indexed with 1 to 10 are high demand users while the remaining users have lower demands. Under demand profile 2, some of the high demand users (indexed with 5 to 10 in demand profile 1) become low demand user while some low demand users (indexed with 11 to 15) become high demand users. Therefore, demand profile 2 (demand at time instant $t+1$) can be treated as time evolution of demand profile 1 (demand at time instant $t$). We can clearly observe that when the demand changes, if we constrain the system not to have any further changes in the user-carrier association, then the demand-supply performance is worse as seen in Fig.~\ref{fig_CADSAT_res4}(c),(d). However, when we relax such constraints, the rate matching performance improves as $Q$ increases. Hence, subplot (b) with $Q=4$ exhibits the best rate-matching along with smaller unmet and unused capacity while subplot (d) with $Q=0$ exhibits the worst performance. However, as we mentioned earlier, with $Q=4$ although we have very good rate-matching, the signalling overhead as well as susceptibility to link outage/degradation with $Q=4$ is much higher than that of $Q<4$ during carrier swapping.
 \begin{table}[h]
\centering
\caption{Impact of $Q$ Values} 
\label{tab-123}
\begin{center}
 \begin{tabular}{||c|| c| c|c |c|c||} 
 \hline
\multirow{2}{*}{Capacity in Mbps}& \multicolumn{5}{|c|}{$Q$ Values} \\ \cline{2-6}
  & 0 & 1 & 2 & 3 & 4 \\ 
 \hline\hline
{\textbf{Unmet}}   & {217.30} & {145.93} & {80.76} & {35.36} & {5.39} \\ 
 \hline
{\textbf{Unused}}  & {212.08} & {132.56} & {68.29} & {23.28} & {0} \\
 \hline
 \hline
\end{tabular}
\end{center}
\end{table}
%\vspace{-3mm}

The unmet and unused capacity values corresponding to different subplots in Fig.~\ref{fig_CADSAT_res4} for different values of $Q$ are provided in Table.~\ref{tab-123}. As mentioned earlier, the unmet and unused capacity gradually improve, i.e., become smaller as we increase the values of $Q$.

\section{Conclusions}
This paper studies the CA scheme in high throughput satellite systems. We propose an efficient multiuser aggregation and access control solution for CA in which the optimal transponder fill-rates and user-carrier association are derived. The performance alalysis of the proposed solution shows that CA can be very useful in enhancing the peak data rate of satellite users as well as in efficiently utilizing the available resources.

Although CA in satellite systems needs to be addressed at different levels of the communication stack, we have limited our focus only to payload level. Physical layer as well as the impact of RF  issues, for example, the impact of spectrum emission musk, spurious emissions, adjacent carrier leakage ration, maximum output power, non-linear satellite channel are left for future works. Furthermore, the synchronization and processing complexity will also be considered in our future CA study.

Furthermore, the CA in this study is limited to GEO satellites, in particular, intra-satellite scenario. The feasibility and performance evaluation of CA in inter-satellite scenario\cite{CA1} as well as in other orbitals, i.e., low earth orbit (LEO), medium earth orbit (MEO) satellite systems is also very important. Note that different CA configurations come up with some inherent advantages and disadvantages over each other. The complexity (at gateway and user terminal level) of implementation of different CA scenarios also vary. The business impact from the satellite operator perspective is also an important issue for CA in satellite systems.

\section*{Acknowledgement}
This work has received funding from the European Space Agency (ESA) funded activity CADSAT: Carrier Aggregation in Satellite Communication Networks. The views of the authors of this paper do not necessarily reflect the views of ESA.

%\vspace{3mm}


\begin{thebibliography}{9}

\bibitem{Sharma}
 S. K. Sharma, S. Chatzinotas and P.-D. Arapoglou, ``Satellite Communications in the 5G Era", IET Digital Library, 2018. 

\bibitem{Vasavada}
Y. Vasavada \textit{et al}., ``Architectures for Next Generation High Throughput Satellite Systems," in \textit{Int. J. Sat. Commun. Net.}, vol. 34, pp. 523-546, 2016. 


\bibitem{Knab2013}
J. J. Knab, ``Optimization of Commercial Satellite Transponders and Terminals,”  in \textit{IEEE Trans. Aerosp. Electron. Syst.}, vol. 49, no. 1, pp. 617--622, Jan. 2013.


\bibitem{Cocco2015}
G. Cocco, T. D. Cola, M. Angelone and Z. Katona,``Radio resource management strategies for DVB-S2 systems operated with flexible satellite payloads” \textit{2016 8th Advanced Satellite Multimedia Systems Conference and the 14th Signal Processing for Space Communications Workshop (ASMS/SPSC),}, Palma de Mallorca, pp. 1--8, 2016.

\bibitem{Knab2015}
J. J. Knab, ``Optimum transponder gain and power for fully loaded satellite,” in \textit{IEEE Trans. Aerosp. Electron. Syst.}, vol. 51, no. 4, pp. 3470--3474, Oct. 2015.

\bibitem{3GPP-CA}
TR 36.808 Evolved Universal Terrestrial Radio Access (E-UTRA); Carrier Aggregation; Base Station (BS) radio transmission and reception.

\bibitem{3GPP-CA1}
 M.~Iwamura, K.~Etemad, M.-H.~ Fong, R.~Nory, and R.~Love, ``Carrier Aggregation Framework in 3GPP LTE-Advanced", in \textit{IEEE Commun. Mag.}, vol. 48, no. 8, pp.60--67, Aug.~2010. 

\bibitem{CA2}
H. Lee, S. Vahid, K. Moessner, ``A survey of radio resource management for spectrum aggregation in LTE-advanced", in \textit{IEEE Commun. Surveys Tuts.}, vol. 16, no. 2, 2nd Quart. 2014, pp. 745--760.

\bibitem{ESA}
European Space Agency (ESA), Carrier Aggregation in Satellite Communication Networks - CADSAT, 2018-2020. https://wwwfr.uni.lu/snt/research/sigcom/projects/cadsat\_carrier\_aggrega\\tion\_in\_satellite\_communication\_networks.

\bibitem{NewTec}
NEWTEC Channel Bonding, [online]. Available: https://www.newtec.eu/technology/channel-bonding.

\bibitem{ZKhan}
Z. Khan, H. Ahmadi, E. Hossain, M. Coupechoux, L. A. Dasilva and J. J. Lehtomäki, "Carrier aggregation/channel bonding in next generation cellular networks: methods and challenges," in \textit{IEEE Network}, vol. 28, no. 6, pp. 34--40, Nov.-Dec. 2014.

\bibitem{DVBS2X}
    ETSI TR 102 376-2 V1.1.1 (2015-11): Implementation guidelines for the second generation system for Broadcasting, Interactive Services, News Gathering and other broadband satellite applications; Part 2: S2 Extensions (DVB-S2X).

\bibitem{GSE}
ETSI TS 102 606 v1.1.1 (2007-10), Digital Video Broadcasting (DVB); Generic Stream Encapsulation (GSE) protocol.


\bibitem{CVX}
M.~Grant and S.~ Boyd. CVX: Matlab Software for Disciplined Convex Programming, version 2.0 beta. http://cvxr.com/cvx, Sep.~2013.

\bibitem{CA1}
R.~Radhakrishnan \textit{et al}., ``Survey of Inter-Satellite Communication for Small Satellite Systems: Physical Layer to Network Layer View," in \textit{IEEE Commun. Surveys Tuts}, vol.~18, no.~2, 4th Quart. 2016, pp.~2442--2473.



\end{thebibliography}
\end{document}